\documentclass[runningheads]{llncs}
\usepackage{graphicx}
\usepackage{booktabs,amsmath}
\usepackage{amssymb}
\usepackage{makecell}
\usepackage{floatrow}
\usepackage{afterpage}

\def\x{{\mathbf x}}

\def\h{{\mathbf h}}
\def\H{{\mathbf H}}

\def\tL{{\tilde L}}

\begin{document}
	\title{Edge-variational Graph Convolutional Networks for Uncertainty-aware Disease Prediction}

	\titlerunning{EV-GCN for Disease Prediction}
	
\author{Yongxiang Huang \and Albert C. S. Chung}

\authorrunning{Y. Huang and A. C. S. Chung}

\institute{
	Lo Kwee-Seong Medical Image Analysis Laboratory,\\
	Department of Computer Science and Engineering,\\
	The Hong Kong University of Science and Technology, Hong Kong, China\\
	\email{\{yhuangch,achung\}@cse.ust.hk}
}
	
	\maketitle            

	\begin{abstract}
		There is a rising need for computational models that can complementarily leverage data of different modalities while investigating associations between subjects for population-based disease analysis. Despite the success of convolutional neural networks in representation learning for imaging data, it is still a very challenging task. In this paper, we propose a generalizable framework that can automatically integrate imaging data with non-imaging data in populations for uncertainty-aware disease prediction. At its core is a learnable adaptive population graph with variational edges, which we mathematically prove that it is optimizable in conjunction with graph convolutional neural networks. To estimate the predictive uncertainty related to the graph topology, we propose the novel concept of Monte-Carlo edge dropout. Experimental results on four databases show that our method can consistently and significantly improve the diagnostic accuracy for Autism spectrum disorder, Alzheimer's disease, and ocular diseases, indicating its generalizability in leveraging multimodal data for computer-aided diagnosis.
	\end{abstract}
	
	\section{Introduction}
	Integrating imaging data with non-imaging data for disease diagnosis is an essential task in clinics. 
	In recent years, the increasing volume of digitalized multimodal data has raised the need for computational models with the capability of exploiting different modalities automatically for improving prediction accuracy and discovering new biomarkers to study the disease mechanism (e.g., Alzheimer's disease) \cite{trojanowski2010update}. 
	Despite the success of convolutional neural networks (CNNs) in medical images \cite{litjens2017survey,huang2019evidence}, exploiting both imaging and non-imaging data in populations in a unified model can be challenging.
	Multimodal learning-based approaches usually summarize features of all modalities with a deep neural network for disease classification \cite{xu2016multimodal}, which ignore the interaction and association between subjects in a population. Graphs provide a natural way to represent the population data and enable the use of powerful tools such as clustering algorithms for disease analysis. Moreover, recent studies on graph convolutional neural networks (GCNs) \cite{wu2019comprehensive,kipf2016semi} have extended the theory of signal processing on graphs \cite{shuman2013emerging} to complement the representation learning limitation of CNNs on irregular graph data.
	
	In this work, we present a generalizable framework to automatically integrate multimodal data in populations for disease prediction with uncertainty estimation. Our contributions include: 
	i) proposing a novel adaptive population graph model for representing multimodal features and associations for subjects and mathematically showing that it can be optimized in conjunction with spectral GCNs, which makes semi-supervised learning with GCNs generalizable for medical databases, 
	ii) proposing Monte-Carlo edge dropout for estimating the predictive uncertainty with respect to the graph topology, which is new and extendable for graph neural networks,  
	iii) designing a well-regularized spectral graph convolutional network for population-based disease prediction, alleviating the over-smoothing problem, and 
	iv) extensively evaluating our method and recent multimodal learning methods on four challenging databases, which shows the proposed method can significantly improve the diagnostic performance for brain analysis and ocular diseases. (To the best of our knowledge, it is also the first study of GCNs on population-based ocular disease analysis.)  
	
	\textbf{Related Work.}  
	Recent studies \cite{kipf2016semi,li2018deeper,wu2019comprehensive} have shown that a graph can serve as a regularizer on node classification tasks for semi-supervised learning with GCNs. However, these methods are evaluated on graph benchmarks where the associations between nodes are inherently defined in the data (e.g., a citation dataset \cite{kipf2016semi}). Contrastively, in the medical domain, the associations between subjects (i.e., nodes) are usually uncertain and multifaceted, especially for multimodal databases. The absence of well-defined relation between two nodes leads to the uncertainty of graph topology, making it hard to adopt GCNs for population-based disease diagnosis. A few recent studies investigated the construction of  affinity graphs by computing the correlation distance between subjects for brain analysis  \cite{kazi2019inceptiongcn,parisot2017spectral}. These methods require to manually tune distance thresholds for different modalities for the graph construction, which can heavily fluctuate the performance, leading to a lack of generalizability.  
	
	\section{Methods}
	
\begin{figure*}[thb]
		\centering
		\centerline{
			\includegraphics[scale=0.35]{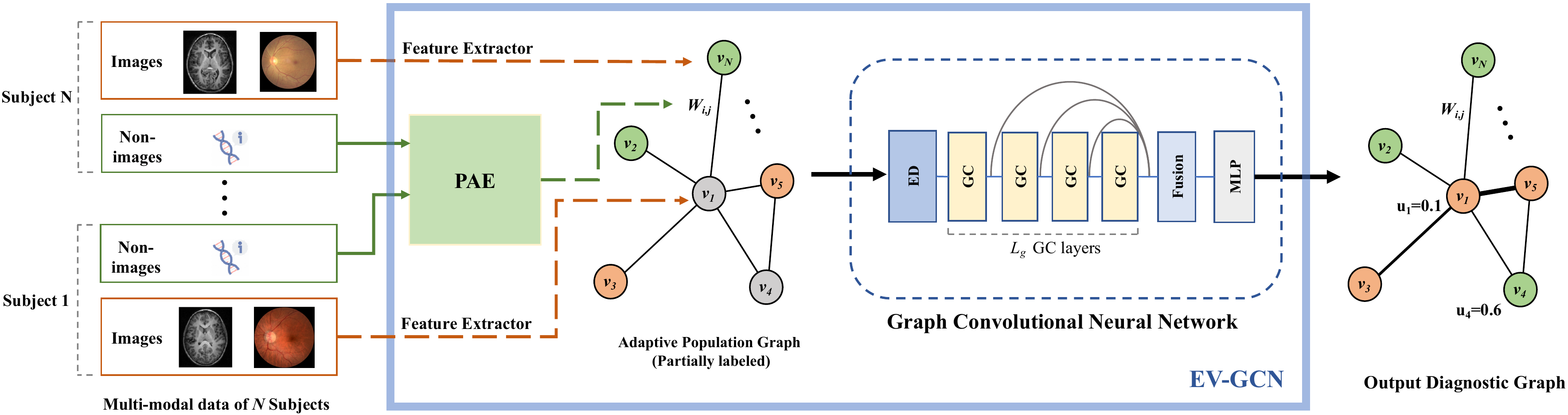}
		}
		\caption{
			Overview of the proposed method. 
			PAE: pairwise association encoder. ED: edge dropout. GC: graph convolution. Fusion: vertex-wise concatenation. Colors in the graphs: green and orange - labeled diagnostic values (e.g., healthy or diseased), grey: unlabeled. $u_i$: predictive uncertainty for subject $i$.}
		\label{fig:fig1}
	\end{figure*}

	In this section, we present the proposed model, namely Edge-Variational GCN (EV-GCN), for incorporating multimodal data  for disease prediction. The overview of the pipeline is depicted in Fig.{~\ref{fig:fig1}}. 
	EV-GCN accepts the imaging features and non-imaging data of $N$ subjects and constructs an adaptive population graph with partially labeled nodes and variational edges, followed by the proposed spectral graph convolutional network with edge dropout for learning to estimate the diagnostic value and the uncertainty for each testing subject.  
	
	\subsection{ Edge-variational Population Graph Modeling }
	Given the observation of $N$ subjects composed of imaging and non-imaging data, let us consider constructing a population graph $G=(V, E, W)$, where  $V$ is a finite set of vertices with $|V|=N$, $E \subseteq V \times V $ is a set of edges with $W$ being the edge weights. To associate a vertex $v \in V$ with the diagnostic features of a subject, we define the node feature $Z_{i} \in \mathbb{R}^{C}$ as a $C$-dimensional feature vector extracted from the imaging data of subject $i$, under the observation that imaging data (e.g., histology images, functional MRI) usually provide the most important evidence for diagnosis. The modeling for the graph connectivity (i.e., edge weight) is critical for the task performance as it encodes the associations between subjects. 
	Unlike previous methods modeling the edge weight statistically \cite{kazi2019inceptiongcn,parisot2017spectral}, we propose to define the edge weight $w_{i,j} \in W$ between the $i$-th and $j$-th vertices as a learnable function of their non-imaging measurements $(\x_i, \x_j)$, considering non-imaging data can provide additional information (e.g., gender, age, gene) complementary to imaging features and explain potential association. 
	The learnable function $f: (\x_i, \x_j) \mapsto \mathbb{R}$ is modeled by the proposed pairwise association encoder (PAE) with trainable parameters $\Omega$ such that $w_{i,j} = f(\x_i, \x_j; \Omega)$.
	
	\subsubsection{Pairwise Association Encoder}  
	The PAE starts by normalizing the multimodal inputs $\x_i$ and  $\x_j$ to zero-mean and unit-variance element-wisely to avoid the gradient vanishing problem, which is important in our setting as data from different modalities have various statistical properties.  After normalization, we use a projection network to map each normalized input to a common latent space $\h_i \in \mathbb{R}^{D_h}$ where cosine similarity can be better applied ($D_h=128$). The projection network is a multi-layer perceptron (MLP) with $L_p$ hidden layers. We set $L_p=1$ in experiments and the latent feature for $\x_i$ is formulated as $ \h_i = \Omega^{(2)} \sigma(\Omega^{(1)} \tilde{\x}_i + b)  $, where $\sigma$ is a ReLU function and $\tilde{\x}_i$ is the normalized input. Formally, the PAE scores the association between vertices $i$ and $j$ as 
	\begin{equation}
	w_{i,j}= \frac{ \h_i ^\top \h_j }{2 \lVert \h_i \rVert  \lVert \h_j \rVert } + 0.5,
	\end{equation}
	which gives the rescaled cosine similarity between two latent features. In training, the parameter $\Omega$ in PAE is initialized by He Initialization. Each hidden layer is equipped with batch normalization and dropout to improve convergence and avoid overfitting. Notably, we find it beneficial and robust to define the pairwise association on the latent space rather than on the input space. 
	
	\subsection{Spectral Graph Convolutions on Adaptive Graphs}
	
	In this section, we first present spectral graph convolutions and prove the connectivity of the proposed adaptive population graph can be optimized in conjunction with a spectral GCN, followed by presenting our GCN architecture. 
	
	A spectral convolution of a graph signal $x$ with a filter $g_\theta$ is defined as  $ g_\theta \star x = U g_\theta U^T x$, where $U$ is the matrix of eigenvectors of the normalized graph Laplacian $L=I_N - D^{-1/2} W D^{1/2} $ with $D$ being the diagonal degree matrix (i.e., $L$ encodes the topological structure of graph $G$).
	The intuition behind is that spatial graph convolutions can be computed in the Fourier domain as multiplications using graph Fourier transform (GFT) \cite{shuman2013emerging}. 
	To reduce the computational cost ($\mathcal{O}(N^2)$), \textit{Chebyshev graph convolution} (ChebyGConv) \cite{defferrard2016convolutional} approximates spectral graph convolution using Chebyshev polynomials. 
	\footnote{ChebyGConv can achieve localized filtering on an irregular weighted graph with moderate computational cost, and comparatively performs well for our tasks. }
	A $K$-order ChebyGConv is given by $g_{\theta'} \star x  \approx  \sum_{k=0}^{K} T_k(\tilde{L}) \theta_k' x  $, where $\tilde{L} $ is the rescaled graph Laplacian, $T_k(\tL) = 2\tL T_{k-1}(\tL)  - T_{k-2}(\tL)$ is the Chebyshev polynomial defined recursively with $T_0(\tL)=1$ and $T_1(\tL)=\tL$, and  $\theta_k'$ are the filter parameters. As an analogy to CNNs, the polynomial term $T_k(\tilde{L})$ acts as a $k$-localized aggregator, i.e., it combines the neighboring nodes that are $k$-step away from the central node, and  $\theta_k'$ acts as a node feature transformer.

	\subsubsection{Claim: } 
	\textit{The graph connectivity modeled by PAE can be optimized in conjunction with a spectral graph convolutional model using gradient descent algorithms. }
	
	\noindent\textit{Proof}:  
	Let us consider a convolution layer $l+1$ in a GCN and its input graph $G^l = (V^l, E^l, W)$ with $|V^l|=N$ 
	$C^l$-dimensional node feature vectors $\mathbf{H}^l \in \mathbf{R}^{N\times C^l}$. The convolution layer computes the output features as
	\begin{equation}
	\label{eq;Eq1}
	\H^{l+1} = \sum_{k=0}^{K} T_k(\tilde{L}) \H^l \mathbf{\Theta}_k^l,
	\end{equation} 
	where $\mathbf{\Theta}_k^l \in \mathbf{R}^{C^l \times C^{l+1}}$  are the filter parameters. 
	Denote $\mathcal{L}$ as the task-related loss function. 
	To optimize the graph connectivity modeled by PAE by gradient descents, we need to guarantee that $\mathcal{L}$ is differentiable w.r.t. the parameters $\Omega$ of PAE. 
	By chain rule, we have $\frac{\partial{\mathcal{L}}}{\partial{\Omega}} = \frac{\partial{\mathcal{L}}}{\partial{\H^l}}  \frac{\partial{\H^l}}{\partial{W}}  \frac{\partial{W}}{\partial{\Omega}} $. Both $\frac{\partial{\mathcal{L}}}{\partial{\H^l}} $ and $\frac{\partial{W}}{\partial{\Omega}} $ are derivable as they correspond to the gradients in the differentiable GCN and PAE respectively. The key is the derivative of the node feature vectors w.r.t. the input edge weights  $\frac{\partial{\H^l}}{\partial{W}} $. For $K=1$, based on Eq. \ref{eq;Eq1}, we can derive $\frac{\partial{\H^l}}{\partial{W}}\big|_{K=1}  = \frac{\partial{(I_N- D^{-1/2} W D^{1/2}) }}{\partial{W}} $, which is derivable.  For $K>1$, since the polynomial term $T_k(\tilde{L}) $ in Eq. \ref{eq;Eq1} is defined recursively for $k>1$,  by expanding $T_k(\tilde{L}) $ and by induction we can derive that $\frac{\partial{\H^l}}{\partial{W}} $  is derivable and not a constant zero. Thus, the graph connectivity can be optimized to minimize the task loss. 

	\subsubsection{GCN Architecture} 
	As depicted in Fig. \ref{fig:fig1}, our GCN model consists of $L_G$ Chebyshev graph convolutional layers, each followed by ReLU activation to increase non-linearity, a fusion block, and an MLP predictor. To alleviate the over-smoothing problem \cite{li2018deeper}
	in deep GCNs, we propose to adopt jumping connections \cite{xu2018representation} with vertex-wise concatenation to fuse the hidden features in each depth, i.e., $\{\H^l\}_{l=1}^{L_G}$.   
	We find jumping connections are more effective than residual connections \cite{he2016deep} in avoiding a performance deterioration as $L_G$ increases. The MLP predictor consists of two $1\times1$ convolutional layers (i.e., vertex-wise transformations) with 256 and $C_k$ channels respectively, followed by a softmax function to drive a $C_k$-class disease probability vector for each subject. We employ cross-entropy loss on the labeled nodes to train the overall model.

	\subsection{Monte-Carlo Edge Dropout for Uncertainty Estimation} 
	
	Motivated by Bayesian approximation for uncertainty estimation in CNNs \cite{kendall2017uncertainties,gal2016dropout,yu2019uncertainty}, we propose \textit{Monte-Carlo edge dropout} (MCED) to estimate the uncertainty for the constructed population graph structure. 
	
	In detail, edge dropout randomly drops a fraction of edges in the graph by placing a Bernoulli distributional mask on each $e \in E$, which can act as a graph data augmenter to reduce overfitting \cite{rong2019truly} and increase the graph sparsity in training. 
	For inference, similar to Monte-Carlo dropout \cite{kendall2017uncertainties} for uncertainty estimation in CNNs, MCED performs $T$ stochastic forward passes on the GCN model under random edge dropout on the population graph, and obtain $T$ disease probability vectors for a subject $i$ :  $\{\textbf{p}_i^{(t)}\}_{t=1}^T$ (We set $T=128$ in our experiments). 
	We adopt the mean predictive entropy \cite{kendall2017uncertainties} as the metric to quantize the uncertainty. Formally, the uncertainty $u_i$ for subject $i$ is given by 
	$ u_i = -\sum_c  \textbf{m}_{i,c} \log \textbf{m}_{i,c} $ 
	and 
	$ \textbf{m}_i = \frac{1}{T} \sum_t \textbf{p}_i^{(t)}$, 
	where $c$ corresponds to the $c$-th class. 
	While MC dropout estimates the uncertainty on the neural network weights \cite{gal2016dropout},
	MCED models the uncertainty on the graph topology, which is orthogonal. 
	
	\section{Experiments \& Results}
	
	In this section, we perform experimental evaluations of the proposed method, comparing our method with several SoTA methods for disease prediction. We consider  two graph-based methods that exploit GCNs (i.e., AIG \cite{kazi2019inceptiongcn} and Parisot et al. \cite{parisot2017spectral}) , a multimodal learning method (i.e., DNN-JFC \cite{xu2016multimodal}), and task-related unimodal methods  \cite{abraham2017deriving,tan2019efficientnet,szegedy2017inception}. 
	
	\subsection{Autism Spectrum Disorder Prediction on the ABIDE database} 
	\subsubsection{Dataset and Experimental Setting} 
	The Autism Brain Imaging Data Exchange (\textbf{ABIDE}) \cite{di2014autism} publicly shares neuroimaging (functional magnetic resonance imaging (fMRI)) and the corresponding phenotypic data (e.g., age, gender, and acquisition site) of 1112 subjects, with binary labels indicating the presence of Autism Spectrum Disorder (ASD). 
	For a fair comparison with the ABIDE state-of-the-art \cite{kazi2019inceptiongcn,abraham2017deriving}, we choose the same 871 subjects consisting of 403 normal and 468 ASD individuals, use 10-fold cross-validation, and perform the same data preprocessing steps \cite{abraham2017deriving}
	to extract a $C=2000$ dimensional feature vector from fMRI representing a subject's functional connectivity \cite{abraham2017deriving}. Phenotypic data is used to compute the pairwise association.
	
	In our experiments, we set Cheybyshev polynomial order $K=3$ and $L_G = 4$. All models are trained using Adam optimizer  with  learning rate $0.01$,  weight decay $5 \times 10^{-5}$, and dropout rate $0.2$ for 300 epochs to avoid overfitting. 
	
	\begin{table}[h] 
		\centering
		\caption{Quantitative comparisons between different methods on ABIDE. MM: multi-modality,  $\times$: only imaging data is used, $\checkmark$: both imaging and non-imaging data are used. $\beta$ is a threshold for constructing a static affinity graph used in AIG.
		} 
		{
			\begin{tabular}{l  c c c c c}
				\toprule
				\textbf{Methods}                                                            & MM                &   Accuracy         & AUC & F1-score  & \#Param.(K)    \\
				\midrule    
				Abraham et al. \cite{abraham2017deriving}    &     $\times$          &       66.80            &    -   &  -   & - \\
				DNN 																&     $\times$      &     $73.25\pm3.69$   & $74.16\pm4.64$  & $74.81\pm4.86$   &  550\\
				DNN-JFC \cite{xu2016multimodal}                &  $\checkmark$   &     $73.59\pm4.15$  & $73.48\pm4.94$    &  $76.89\pm4.27$	& 635 \\
				Parisot et al. \cite{parisot2017spectral}     		& $\checkmark$   &   $75.66\pm4.69$        &   $81.05\pm6.13$  &  $78.85\pm4.66$  &   96 \\
				AIG \cite{kazi2019inceptiongcn}       & $\checkmark$      &$76.12\pm6.83$       & $80.11\pm6.49$   &  $79.27\pm5.27$   &    290 \\
				AIG, $\beta=3$  \cite{kazi2019inceptiongcn}    & $\checkmark$   &   $72.10\pm7.12$    &$75.43\pm8.85$     & $73.55\pm6.94$     & 290 \\
				\midrule    
				This work: 																								&         &           &        &   & \\
				EV-GCN 										& $\checkmark$     &  $80.83\pm4.92$      & $\mathbf{84.98\pm5.74}$	&  $81.24\pm5.76$ &  133\\   
				+ MCED										& $\checkmark$       &  $\mathbf{81.06\pm4.83}$      & $84.72\pm6.27$  &  $\mathbf{82.86\pm5.51}$  &  133\\
				\bottomrule
			\end{tabular}
			\label{tab:table1}
		}
	\end{table}

	\begin{table}[h] 
		\centering
		\caption{Ablation study for this work. This table shows how different factors affect the performance of our method. 
			The compared GCN architectures share the same depth. Plain: sequential GC layers. Inception: InceptionGCN  \cite{kazi2019inceptiongcn}.
		}  
		{
			\begin{tabular}{  l | c c c | c c  c| c }
				\hline
				Factor          & \multicolumn{3}{c|}{Graph Construction}     &   \multicolumn{3}{c|}{GCN Architecture}	&  \multicolumn{1}{c}{Edge Dropout}   \\
				\hline 
				Method			&   Random  & Affinity  & Adaptive  &  Plain   & Inception &  Ours        & w/o ED	 \\
				\hline 
				Accuracy  	   &    65.67	  &   76.81 &  \textbf{80.83}      &  79.79 & 80.02	 &		80.83     		&	80.58  	\\
				AUC  		     &    73.97		&   80.49  &  \textbf{84.98}       &   83.10   & 83.74   	&  84.98 	&   84.27 		\\
				\hline 
			\end{tabular}
			\label{tab:table2}
		}
	\end{table}   
	
	\subsubsection{Results and Analysis}  
	Table \ref{tab:table1} shows the comparative results on ABIDE, where the mean and confidence interval ($p < 0.05$) are computed across ten different initialisation seeds. 
	DNN-JFC \cite{xu2016multimodal} summarizes the features of all modalities by jointly fully connected layers, which marginally outperforms a DNN (i.e. MLP) on the fMRI features. Comparatively, graph-based methods (Parisot, AIG, and ours) yield larger performance gains, benefiting from exploiting associations between subjects in the population graphs. The proposed method, EV-GCN+MCED, obtains an average accuracy of $81.06\%$, outperforming the recent SoTA method AIG \cite{kazi2019inceptiongcn}, which employs static affinity graphs with InceptionGCN, by a margin of $4.94\%$ accuracy with fewer parameters. We notice that the performance of AIG is highly sensitive to the threshold $\beta$ for computing age affinity, where the best $\beta$=2 yields an average accuracy of  $76.12\%$. 
	To investigate the importance of learning an adaptive graph with variational edges, we train our GCN architecture on the same affinity graph used in AIG \cite{kazi2019inceptiongcn} and on a population graph with random connections. As depicted in Table \ref{tab:table2}, it results in a $4.02\%$ (Affinity) and $15.16\%$ (Random) accuracy drop respectively, indicating that the adaptive graph modeling is indeed key to achieving the best possible performance. Meanwhile, the effectiveness of the proposed GCN architecture and edge dropout regularization in training  are ablatively validated in Table \ref{tab:table2}. 
	
	\begin{table}[h]
		\centering
		\floatbox[{\capbeside\thisfloatsetup{capbesideposition={left,top} }}]{table}[\FBwidth]
		{\caption{Accuracy and uncertainty for models with different association (i.e. edge) inputs.  We show the  uncertainty as the mean value of all test subjects. 
		}}
		{
			\begin{tabular}{l  c c }
				\toprule
				Association input 									  & 	Accuracy    &   Uncertainty      \\
				\midrule  
				Random Noise            					   			 &  		  65.67		  &     0.620 		 \\
				Gender          					  		  	 &    	  78.53		 &        0.465   \\ 
				Gender, Age          					  		&    	78.07	  	&    0.394       \\
				Site, Gender, Age          			  &    	 \textbf{81.06}		  &    	\textbf{0.307}      \\
				\bottomrule
			\end{tabular}
			\label{tab:tableUnc}
		}
	\end{table}

	To analyze what the uncertainty estimated by MCED captures, we give the accuracy and uncertainty for models trained with different association sources (i.e. input for PAE) in Table \ref{tab:tableUnc}. The results demonstrate that the graph uncertainty, which is approximated by averaging the predictive uncertainty of all test subjects, can be gradually eliminated with sufficient information for learning the pairwise association. 
	The results in Table \ref{tab:table1} and Table \ref{tab:tableUnc} show that MCED uncertainty estimation, on one hand, can be adopted as an ensemble approach to improve the diagnostic performance (+$1.62\%$ F1-score), and on the other hand, can be used to detect patients with highly uncertain diagnostic value, which is important for safety-critical CAD systems in reducing misdiagnosis rate. 
	
	\subsection{Alzheimer's Disease Prediction on ADNI} 
	\textbf{ADNI} \cite{adni_url} 
	is a large-scale database and contains longitudinal brain MRI, PET data, genetic, and phenotypic information of over 1700 adults for Alzheimer's disease (AD) study. 
	In this work, we select the same 1675 samples with Mild Cognitive Impairments (MCI) used in Parisot \cite{parisot2017spectral} to facilitate a fair comparison, among which 843 acquisitions will convert to AD as diagnosed during the follow-up.
	For the imaging feature vector, we use the volumes of $C=138$ segmented brain structures extracted from MRI using MALP-EM \cite{ledig2015robust}, which are proven effective biomarkers for AD assessment \cite{ries2008magnetic}. We use phenotypic (age, gender) and genetic (APOE \cite{mosconi2004mci}) data for computing the pairwise association. 
	\subsubsection{TADPOLE} \cite{marinescu2018tadpole}  is a preprocessed subset of ADNI, consisting of 557 subjects each with over 350 multimodal features. Following AIG \cite{kazi2019inceptiongcn}, the task is to classify each subject into three classes: cognitive normal, MCI, and AD. 
	We use the segmentation features derived from MRI and PET data to obtain a 340-dimensional feature vector for each subject and use the phenotypic data, APOE and FDG-PET biomarkers for the graph construction. 
	
	\begin{figure}[h]
		\centering
		\begin{tabular}{ccc}
			\includegraphics[scale=0.25]{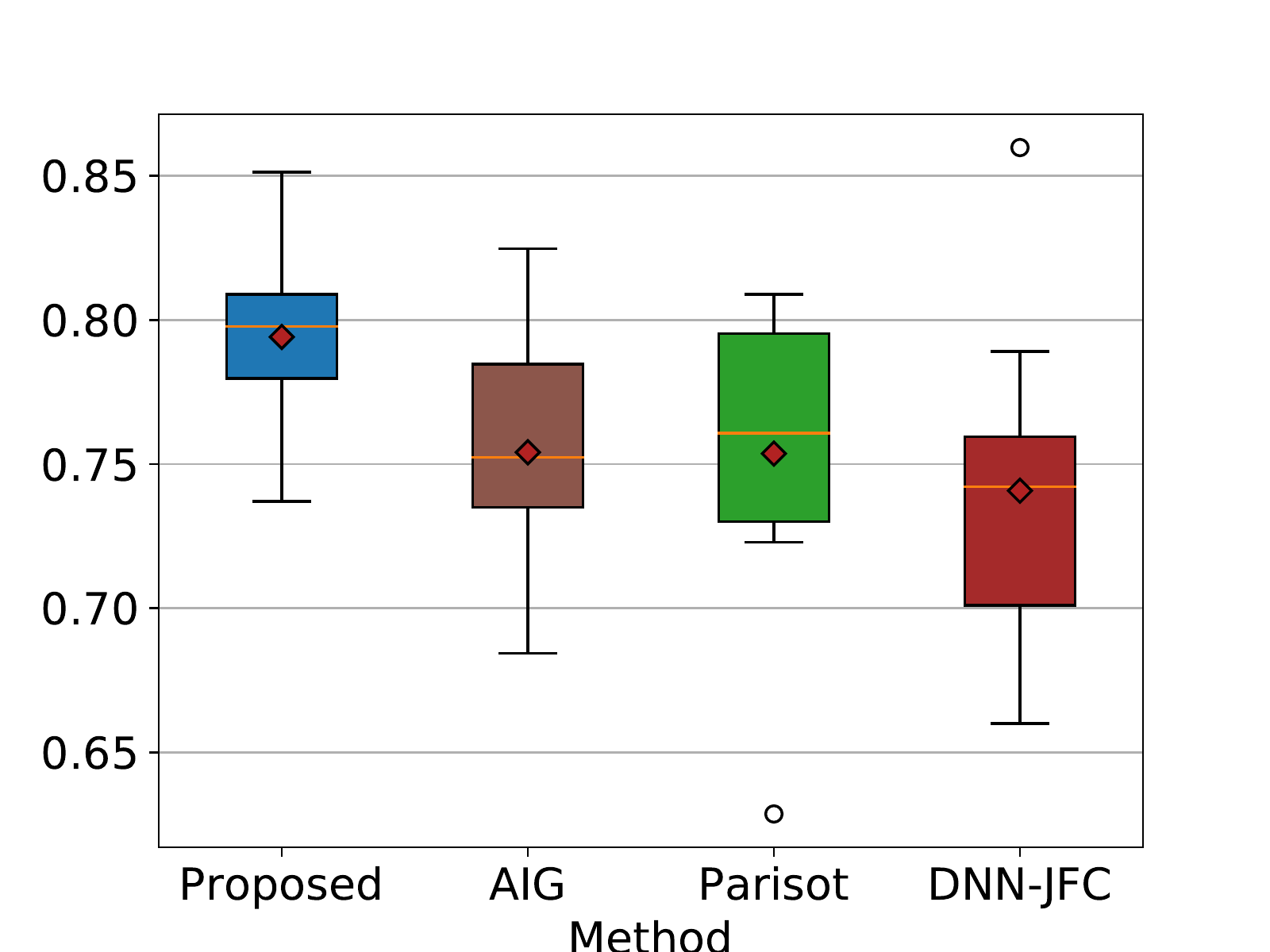} & 
			\includegraphics[scale=0.25]{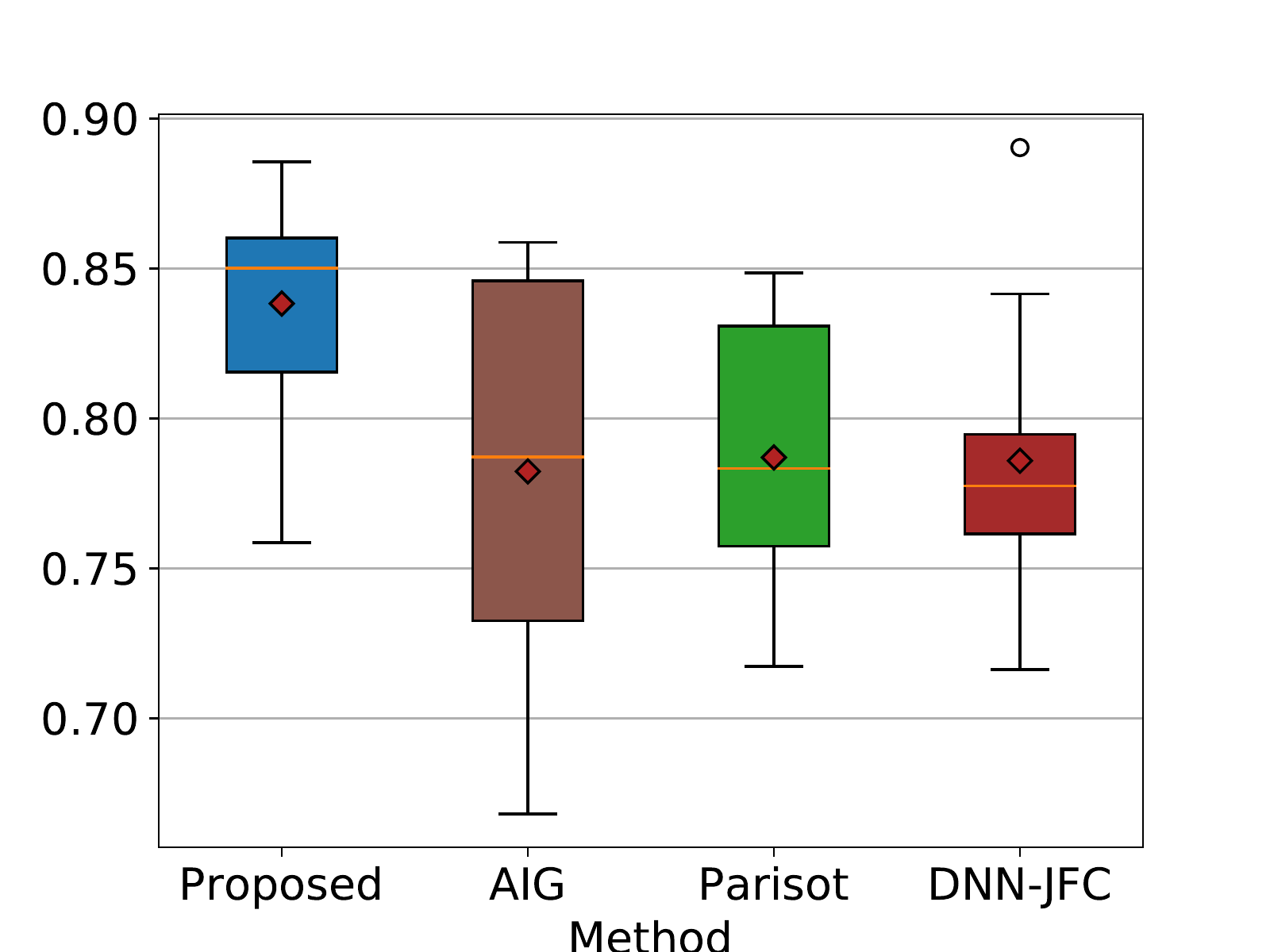} & 
			\includegraphics[scale=0.25]{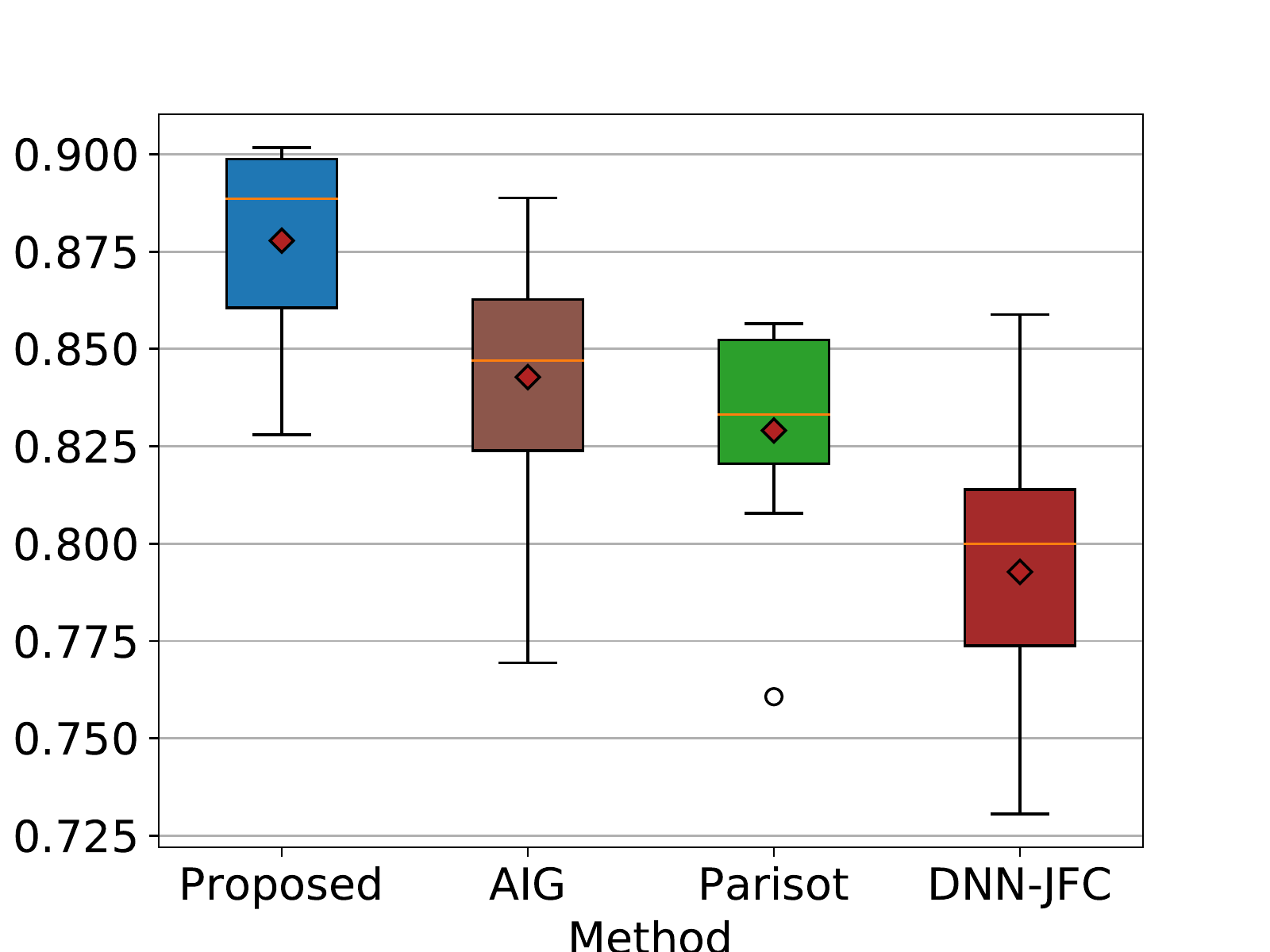}  \\
			(a) ADNI Accuracy   & (b) ADNI AUC & (c)  TADPOLE Accuracy
		\end{tabular}
		\caption{ Comparative boxplots on ADNI (a,b) and TADPOLE (c) for Alzheimer's disease prediction. Results are computed from 10-fold cross-validation.
		} 
		\label{fig:figAD}
	\end{figure}

	\subsubsection{Results} 
	Comparative boxplots for ten folds between the four methods are shown in Fig. \ref{fig:figAD}. We can observe that the proposed method (EV-GCN+MCED) outperforms the competing methods on both datasets. For prediction AD conversion on ADNI, we achieve an average accuracy of $79.4\%$, corresponding to a $3.9\%$ increase over the competing method Parisot \cite{parisot2017spectral}.  For TADPOLE, our method obtains an average accuracy of $87.8\%$, outperforming the recent SoTA method AIG  \cite{kazi2019inceptiongcn} ($84.3\%$). The results also imply the generalizability of our method. 
	
	\subsection{Ocular Disease Diagnosis on the ODIR dataset}
	\subsubsection{Dataset and Experimental Setting} 
	The ODIR dataset \cite{odri_url} shares fundus photographs and non-imaging data including age, gender and diagnostic words of 3000 patients. Each patient has 8 binary labels for 7 types of ocular diseases including diabetes, glaucoma, etc. We compare our method with two recent SoTA CNNs \cite{tan2019efficientnet,szegedy2017inception} and the graph-based SoTA AIG \cite{kazi2019inceptiongcn}, using 5-fold cross-validation. For population graph construction, we use the compared CNN without classification layers as a feature extractor to derive a $C$ dimensional feature vector from the fundus images (for both eyes) of a patient ($C=3072$ for InceptionV4 \cite{szegedy2017inception} and $C=2048$ for EfficientNet-B0 \cite{tan2019efficientnet}), and use the demographic data for association modeling. Diagnostic words are not used  to avoid label leaking. 
	
	\begin{table}[h]
		\centering
		\floatbox[{\capbeside\thisfloatsetup{capbesideposition={left,top} }}]{table}[\FBwidth]{
			\caption{Quantitative results on ODIR. (I) or (E): InceptionV4 \cite{szegedy2017inception} or EfficientNet \cite{tan2019efficientnet} is used for imaging feature extraction.  D: Diabetes, G: Glaucoma, C: Cataract, Overall: all 8 classes.} 
		}
		{
			\begin{tabular}{l  c c c c}
				\toprule
				Methods                            &   D         & G     & C & Overall \\
				\midrule    
				InceptionV4 \cite{szegedy2017inception}     &       64.26         &    69.89        &   95.34 &  $84.00\pm11.35$ \\
				EV-GCN (I)                              &    68.49          &  70.31      & 94.22            & $86.62\pm11.26$ \\
				EfficientNet \cite{tan2019efficientnet}             &      66.90          &    71.91           & \textbf{95.91} &  $84.31\pm12.11$ \\
				EV-GCN (E)                              &     \textbf{70.78}         & \textbf{73.24} & 95.18    &
				$\mathbf{87.63\pm9.88}$ \\
				AIG \cite{kazi2019inceptiongcn} (E)                 & 58.61&    68.08    & 87.60 &  $78.47\pm14.79$ \\
				\bottomrule
			\end{tabular}
			\label{tab:tableODIR}
		}
	\end{table}

	Table \ref{tab:tableODIR} shows the comparative results in terms of AUC for different types of ocular diseases. 
	We see that the EV-GCN can improve the classification performance for both EfficientNet and InceptionV4 (i.e., InceptionResNet-V2 \cite{szegedy2017inception}) on ODIR, e.g. $4\%$ improvement for Diabetic detection. On average, the proposed method can improve the performance of a pre-trained SoTA CNN by $2.97\%$ for fundus image classification, by learning to incorporate the complementary non-imaging data encoded in the graph. It is interesting to note that the static graph-based method \cite{kazi2019inceptiongcn}, where the required thresholds are already finetuned, degrades the performance for EfficientNet, which reassures the robustness of constructing a learnable population graph compared to a hand-crafted one.
	
	\section{Discussion \& Conclusions}
	In this paper, we have proposed a generalizable graph-convolutional framework to tackle the challenges in learning from multi-modal data for disease prediction. Unlike previous methods, the proposed method does not hand-engineer a similarity population graph but learn to construct the graph connectivity which is mathematically proven to be optimizable with GCNs. 
	The proposed Monte-Carlo edge dropout is the first study on graph uncertainty estimation for GCNs and is experimentally validated to be beneficial, while we admit that it requires further theoretical justification in future work.  
	Extensive experimental results show that the proposed method can achieve superior performance on brain analysis and ocular disease prediction. Additionally, the estimated predictive uncertainty allows detecting the uncertain samples for clinical intervention, contributing to a safer deep learning-assisted diagnosis system. 
	We believe such an extendable method can have a great impact in unlocking a better use of multi-modal data in populations for computer-aided diagnosis in clinics.
	
	
	\bibliographystyle{splncs04}
	
	\bibliography{refs}
	
\end{document}